\documentstyle[12pt,amsfonts,epsfig]{elsart}
\begin{document}
\def\O{{\mathcal O}}
\def\Z{{\mathcal Z}}
\def\L{{\mathcal L}}
\def\D{{\mathcal D}}
\newcommand\A{\mbox{\boldmath $A$}}
\newcommand\x{\mbox{\boldmath $x$}}
\newcommand\y{\mbox{\boldmath $y$}}
\newcommand\nab{\mbox{\boldmath $\nabla$}}
\newcommand\dd{{\rm d}}
\newcommand\I{{\rm i}}
\def\beq{\begin{equation}}
\def\eeq{\end{equation}}
\def\bea{\begin{eqnarray}}
\def\eea{\end{eqnarray}}
\begin{flushright}
HIP-1998-15/TH\\
\end{flushright}
\begin{frontmatter}
\title{Vortices and the Ginzburg-Landau phase transition\thanksref{thank0}}
\author{A.~Rajantie\thanksref{thank1}}
\address{Helsinki Institute of Physics\\P.O. Box 9\\
FIN-00014 University of Helsinki, Finland}
\thanks[thank0]{Talk given in Symposium on Quantum Phenomena at Low 
Temperatures, Lammi, Finland, 7--11 January, 1998.}
\thanks[thank1]{e-mail: arttu.rajantie@helsinki.fi}
\begin{abstract}
The methods for studying the role of vortex loops in
the phase transition of the Ginzburg-Landau theory of superconductivity
using lattice Monte Carlo simulations are discussed. 
Gauge-invariant observables that measure the properties
of the vortex loop distribution are defined.
The exact relations
between the lattice and continuum quantities make it possible
to extrapolate the results of the simulations to the continuum limit.
The relationship between spontaneous symmetry breaking and
phase transitions is also reviewed, with an emphasis on the fact
that a local symmetry cannot be broken.
\end{abstract}
\begin{keyword}
Superconductivity, Vortices, Lattice simulations
\end{keyword}
\end{frontmatter}

\section{Introduction}
Vortex lines result from the classical Ginzburg-Landau equations
of motion, when the behavior of a Type II superconductor
in the presence of an external magnetic field is calculated
\cite{ref:nielsenolesen}.
However, in the classical approximation the effect
of thermal fluctuations is neglected, and near the critical temperature
this is not justified anymore \cite{ref:blatter}. 

The inclusion of the fluctuations means moving from the Ginzburg-Landau
equations to a three-dimensional field theory called the 
Ginzburg-Landau theory or the U(1)+Higgs model.
One striking consequence of the fluctuations is that they restore the
gauge symmetry at all temperatures \cite{ref:elitzur}. Therefore,
the phase transition cannot be understood to be a consequence
of spontaneous symmetry breakdown.
One property that is expected to distinguish between the phases
is the behavior of the vortex loops created by the thermal fluctuations
\cite{ref:Chavel}.

No reliable methods are available to study the behavior of the
vortices directly from the continuum theory. Approximative
approaches have been used both in analytical  
and in numerical 
calculations \cite{ref:Chavel,ref:kleinert,ref:vortexpapers}. 
However, the only way to be able to systematically
remove the error of the approximation is to perform Monte Carlo simulations
with a discretized lattice version of the full theory and then extrapolate
the results to zero lattice spacing. Such simulations have been performed
and the results were reported in Ref.~\cite{ref:PRL}, from the point of
view of particle physics. The purpose of this
talk is to discuss the starting points and the ideas of these simulations
and to make a link to condensed matter physics.

This talk is organized as follows. Section 2 contains the
starting points of our analysis: the basic properties of
the continuum Ginzburg-Landau theory. 
In Section 3, we review the phenomenon of
spontaneous symmetry breaking and its absence in gauge theories.
The definition of the lattice version of the Ginzburg-Landau
theory is given in Section 4, and its vortex configurations are discussed
in Section 5, 
where we also discuss the observables measured
in Monte Carlo simulations. 

\section{Ginzburg-Landau theory of superconductivity}
The Ginzburg-Landau theory contains two fields, a complex scalar field
$\psi$ and a real vector-valued gauge field $\A$.
The action of the
theory (the Ginzburg-Landau energy) is
\beq
\label{equ:contlagr}
S=\int\dd^3x\left[
\left|\left(\nab-\I
\A\right)\psi\right|^2
+y\left|\psi\right|^2
+x\left|\psi\right|^4
+\frac{1}{2}\left(\nab\times\A\right)^2\right].
\eeq
The theory is parameterized by two dimensionless variables, $x$ and $y$.
The fields and the coordinate $\x$ have been scaled to dimensionless 
quantities. The parameters $x$ and $y$ are uniquely fixed by the
requirement that the gauge coupling constant is scaled to unity.
It is a trivial matter
to transform the results to physical dimensions, if desired.
The values of the parameters $x$ and $y$ can be derived from
the microscopical BCS theory of superconductivity:
\beq
y=\frac{1}{q^4}\left(\frac{T}{T_0}-1\right),\quad 
x=\frac{g}{q^2},
\eeq
where
$g\approx 111.084\left(T_0/T_F\right)^2\sim 10^{-6}$, and
$q\approx 0.730\sqrt{e^2v_F/\hbar c^2}\sim 10^{-2}$ 
\cite{ref:kleinert,ref:U1sim}.
It can be seen that $x=\kappa^2$ 
depends on the material of the superconductor and is small
in Type I superconductors and large in Type II superconductors.
The parameter
$y$ depends on the temperature. Since the factor multiplying the
temperature in $y$ is so large, $1/q^4\sim 10^8$, the temperature range
in which $y$ is small is very narrow.

The Lagrangian (\ref{equ:contlagr}) is invariant under (local)
gauge transformations
\beq
\label{equ:gauge}
\psi(\x)\rightarrow\exp\left(\I\theta(\x)\right)\psi(\x),\quad
\A(\x)\rightarrow\A(\x)-\nab\theta(\x),
\eeq
where $\theta(\x)$ is an arbitrary real-valued function. In more 
complicated theories, $\theta(\x)$ can be an element of some non-Abelian
Lie group.

When $T$ is far enough from the critical temperature, the fluctuations
are often assumed to be negligible. Then one can use the mean-field approach
and derive the
Ginzburg-Landau equations as the
Euler-Lagrange equations of motion from the action (\ref{equ:contlagr})
and solve them to find out the lowest-energy field configuration
$(\psi_{EL}(\x),\A_{EL}(\x))$.
When $T>T_0$, i.e.~when $y>0$, the solution is trivial and both fields vanish.
At temperatures below $T_0$
the system appears to be in the broken phase, in which the
field $\psi$ has a non-zero value. Because of this, the Higgs mechanism
\cite{ref:Higgs} gives the photon a non-zero mass and
explains
the Meissner effect. Since the solution 
$(\psi_{EL}(\x),\A_{EL}(\x))$ is not invariant
under the gauge transformation (\ref{equ:gauge}),
the state of the system seems to break the symmetry spontaneously.
This is then interpreted to imply a second-order phase transition,
analogously to ferromagnets. A second-order transition is indeed what
seems to be observed in experiments, but there are still fundamental
difficulties in this interpretation.

\section{Spontaneous symmetry breaking}
In statistical mechanics or quantum field theory, the state of the system 
is specified by the observables, i.e.~the 
expectation values of different functions $\langle f[\psi]\rangle$
\cite{ref:lebellac}.
In thermal equilibrium the state is given by the Gibbs measure:
\beq
\label{equ:expO}
\langle f[\psi]\rangle=Z^{-1}\sum_\psi f[\psi]\exp\left(-\beta H[\psi]\right),
\eeq
where the sum is taken over all the possible configurations of the 
system\footnote{In 
quantum statistical mechanics one should take a trace in the 
Fock space instead
of a sum, but the trace can be transformed to a similar form with $d+1$
dimensions \cite{ref:feynmanhibbs}, 
and therefore we will not consider it separately.}
and $Z$ is a constant defined by $\langle 1\rangle=1$.
The definition (\ref{equ:expO}) can be extended to the infinite case by 
defining the system in a box of size $L^d$, where $d$ is the dimensionality
of the system, and taking $L$ to infinity. In the infinite-volume limit
the analyticity of the observables is no longer guaranteed. The
possible non-analyticities are called phase transitions.

Let us consider a model defined on an $L^d$ lattice and 
given by a Hamiltonian $H_{L,h}[\psi]$, 
where $\psi$ denotes the fields of
the theory and $H_{L,h}$ is a linear function of the real variable $h$ 
(typically an external field coupling to $\psi$). 
Suppose that when $h=0$ the Hamiltonian is invariant under
some transformation $\Lambda$ of the fields, 
i.e.~$H^\Lambda_{L,0}[\psi]\equiv H_{L,0}[\Lambda\psi]=H_{L,0}[\psi]$.
Now let us choose some local
function of the fields $f[\psi]$, 
i.e.~one that depends only on the values of $\psi$ at finitely many points.
For any values of $h$ and $L$, we will denote the expectation value of 
$f[\psi]$ by
\beq
\label{equ:latexp}
\langle f[\psi]\rangle_{L,h}=Z^{-1}_{L,h}\sum f[\psi]
\exp\left[-\beta H_{L,h}\right].
\eeq
If, for any such $f[\psi]$,
\beq
\label{equ:SSB}
\langle f[\psi]\rangle\equiv
\lim_{h\rightarrow 0_+}\lim_{L\rightarrow\infty}\langle f[\psi]\rangle_{L,h}
\eeq
is non-invariant, 
i.e.~$\langle f[\Lambda\psi]\rangle\neq\langle f[\psi]\rangle$
the symmetry is said to be spontaneously broken in the infinite-volume
state. This implies that the equilibrium state is non-unique, since
it is related by a symmetry transformation to another state, which
can be obtained by using $H^\Lambda_{L,h}$ instead of 
$H_{L,h}$ in Eq.~(\ref{equ:latexp}).
Thus $\langle f[\psi]\rangle$ is discontinuous at $h=0$,
i.e.~the system is at a first-order transition line. If the symmetry is broken
only at low temperatures, there is some critical temperature $T_c$,
at which the first-order line terminates.
At that point the system undergoes a first- or second-order transition 
to the symmetric phase.

In many models, e.g.~in the Ising model \cite{ref:ising},
the phase transition occurs because the symmetry is broken at
low temperatures. Nevertheless,
one can show that one-dimensional systems with
only local interactions cannot have spontaneous symmetry breakdown.
A less trivial result is the Mermin-Wagner theorem \cite{ref:merwag}, 
which states that 
in two dimensions a continuous symmetry cannot be broken. From 
our point of view the most interesting
such result was proven by Elitzur \cite{ref:elitzur}: A local symmetry
cannot be broken. This means that the equilibrium state of the
theory is gauge invariant at all temperatures. 
There can certainly be phase transitions in the theory, but one needs
more complicated arguments than spontaneous symmetry breaking to explain
them. For example in the confinement transition of gauge field theories
topological properties of the theory are expected to play an important
role. Similarly, in the two-dimensional XY model 
condensation of vortices causes a Korterlitz-Thouless
transition although
spontaneous symmetry breaking is forbidden by the Mermin-Wagner theorem.
On the other hand in the SU(2)+Higgs theory the symmetry-breaking
phase transition predicted by the mean-field theory disappears 
completely in the non-perturbative regime of the parameter space when
fluctuations are taken into account \cite{ref:SU2higgs}.

Despite the Elitzur's theorem, arguments based on broken symmetry
in gauge theories work often very well. The reason is that one 
fixes the gauge to get
rid of the unphysical degrees of freedom. Without gauge fixing,
the expectation value of any non-invariant
function coincides with the expectation value of its average on
the gauge orbit.
For example, the expectation value $\langle\psi\rangle$ of the order
parameter field vanishes, since $\psi$ transforms covariantly
under gauge transformations (\ref{equ:gauge}).

Gauge fixing breaks the gauge symmetry
explicitly. Gauge-invariant quantities remain unchanged
but non-invariant quantities can acquire non-trivial values.
For example in the Landau gauge it is possible that $\langle\psi\rangle$
is non-zero, and it actually acts as an order parameter, obtaining 
non-zero values for some
parameter values while vanishing for others \cite{ref:Kennedy}.
However, it is not a local order parameter in the sense of
Eq.~(\ref{equ:SSB}). Namely, in the Landau gauge $\nab\cdot\A=0$, and
therefore
\beq
\label{equ:landau}
\psi(\x)\exp\left[
\frac{\I}{4\pi}\int\dd^3y\frac{\A(\y)\cdot(\y-\x)}{|\y-\x|^3}\right]=
\psi(\x).
\eeq
Since the left-hand side of the equation is gauge invariant, it is
really its expectation value we are calculating, but it is a non-local
quantity.
We can see from Eq.~(\ref{equ:landau}) 
that a non-zero expectation value of $\psi$ in the Landau
gauge implies that the invariance under transformations (\ref{equ:gauge})
with $\theta$ constant in space is broken, since in that case the exponential 
is unchanged in the transformation while $\psi$ transforms as usual.

Since gauge-invariant quantities are unchanged in the gauge fixing,
one can use perturbation theory in the gauge-fixed theory to calculate
their expectation values, and the resulting expansion is the correct
one \cite{ref:frohlich}. This explains why the arguments based on
symmetry breaking work so well: The order parameter may have a
non-zero value in the particular gauge used in the calculation.
Nevertheless, this does not guarantee a phase transition in the
original theory without gauge fixing.

\section{Lattice model}
In our case the analogue of $\beta H$ is the action $S$ given in 
Eq.~(\ref{equ:contlagr}). However, it is non-trivial to define the sum in
Eq.~(\ref{equ:expO}) in a continuum theory.
In a perturbative approach it can be done with some standard renormalization
method, and we interpret the action (\ref{equ:contlagr}) as renormalized
in the $\overline{\rm MS}$ scheme. 
In Monte Carlo simulations the problem is solved by defining the theory
on a lattice with lattice spacing $a$ and then
taking the limit $a\rightarrow 0$. In our case this can be
done consistently \cite{ref:contU1}.

The lattice 
action\footnote{We 
use the non-compact formulation, i.e.~the actual gauge group
is $\mathbb R$. In the compact formulation, the phases are analytically
connected and the vortex lines are not closed loops.
Both formulations are expected to have the same continuum limit, 
Eq.~(\ref{equ:contlagr}).}
is given by the Lagrangian
\beq
\label{equ:latlagr}
\L =\frac{1}{2a}\sum_{i<j}\alpha_{ij}^2(\x)
\!+\!\frac{1}{2}\sum_i\left|\psi(\x)-U_i(\x)\psi(\x+\hat\imath)\right|^2
\!+\!\lambda\left(|\psi(\x)|^2-v^2\right)^2,
\eeq
where
$\alpha_{ij}(\x)=\alpha_i(\x)+\alpha_j(\x+\hat\imath)-
\alpha_i(\x+\hat\jmath)-\alpha_j(\x)$ and $U_i(\x)=\exp[\I\alpha_i(\x)]$.
Now $\alpha_i(\x)$ is a real number defined on each link between the lattice
sites and corresponds to the continuum gauge field $\A(\x)$, and
$\psi(\x)=\rho(\x)\exp[\I\gamma(\x)]$ is a complex field defined on the sites.
The gauge transformation (\ref{equ:gauge}) becomes
\beq
\gamma(\x)  \to  [\gamma(\x)+\theta(\x)]_\pi,\quad
\alpha_i(\x) \to 
\alpha_i(\x)+\theta(\x)-\theta(\x+\hat{\imath}), \label{equ:gaugelat}
\eeq
where $\theta(\x)$ is a real valued function on the lattice,
and the notation $[X]_\pi$ means $X+2\pi n\in(-\pi,\pi]$ with $n\in
{\mathbb Z}$.

The continuum Ginzburg-Landau theory is obtained by taking the limit
$a\rightarrow 0$ in the lattice theory in such a way that the long-range
properties remain unchanged. This means that we have to vary the values
of the lattice couplings $v^2$ and $\lambda$ as functions of 
$a$ \cite{ref:parisi}. In
our case the leading terms in $a$ can be obtained in perturbation
theory by calculating some physical correlator both in lattice and in
continuum perturbation theory and requiring that they coincide.
The required two-loop calculation was performed in Ref.~\cite{ref:oma} 
and it gives us
the lattice 
parameters $\lambda$, $v^2$ as functions of $x$, $y$ and $a$:
\bea
\label{equ:latren}
\lambda&=&\frac{1}{4}xa,\nonumber\\
v^2&=&-\frac{a}{x}\left[
y-{3.1759115(1+2x)\over2\pi a}\right.\nonumber\\
&&\left.\phantom{-\frac{a}{x}[}
-\frac{
(-4+8x-8x^2)(\log(6/a)+0.09)-1.1+4.6x}{16\pi^2}\right].
\eea
The result is exact in the limit $a\rightarrow 0$.
Clearly, $\lambda$ vanishes in the
continuum limit. Furthermore, the coefficient of the term $\alpha_{ij}^2$
diverges yielding $\alpha_{ij}=0$. Therefore the short-range properties
of the theory in the continuum limit are given by a massless free-field
theory. However, at large distances the behavior is different,
since then also
higher corrections in $a$ yield a non-vanishing contribution.

The phase structure of the lattice U(1)+Higgs model is relatively well 
understood \cite{ref:U1phases}. There are two phases,
the normal phase with a massless photon and the superconducting phase
with a massive photon, i.e.~the Meissner effect.
The system has been rigorously proven to be in the normal phase
if $a$ is larger than some
constant $a_0>0$, or
if $v^2$ is smaller than some $a$-independent value $v^2_0(\lambda)\ge 0$.
However, at values of $a$ smaller than some constant $a_1>0$ and $\lambda$
larger than some $\lambda_1>0$, the
system is in the superconducting phase provided $av^2>a_1v_1^2$ with
some constant $v_1^2$. Also, if $\lambda v^2$ and $av^2$ are positive,
the system will be in the superconducting phase for small enough $\lambda$.
However, the values of the constants appearing in these statements
are not known, and therefore the rigorous results do not tell much about
the continuum limit $a\rightarrow 0$ of the theory.

\begin{figure}
\begin{center}
\epsfig{file=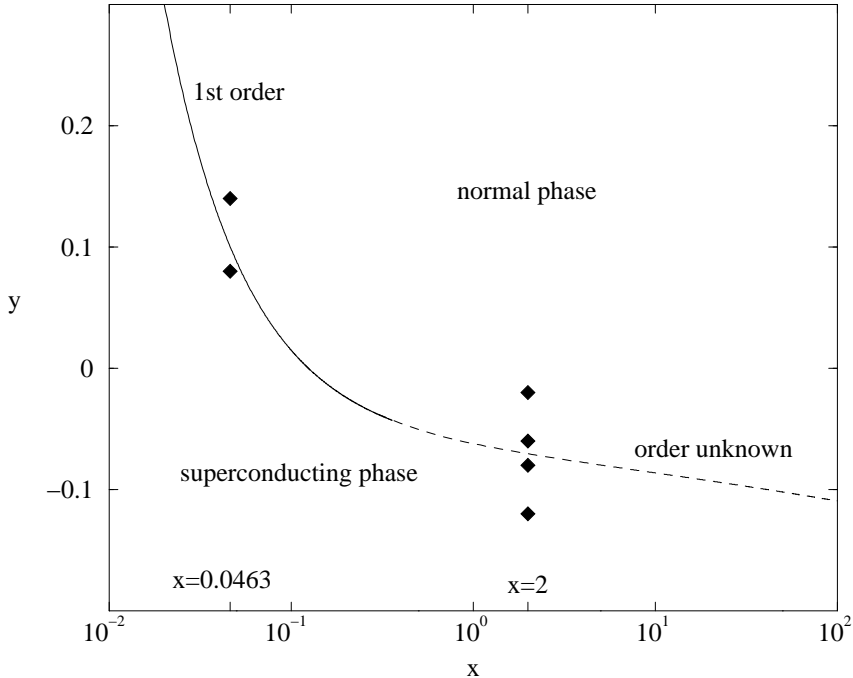,width=13cm}
\end{center}
\caption{
\label{fig:phasediag}
The phase diagram of the continuum theory \cite{ref:U1sim}. 
The diamonds show the
parameter values at which the vortex simulations have been performed.}
\end{figure}

Results of numerical Monte Carlo simulations (see Ref.~\cite{ref:U1sim}
and references therein)
can be extrapolated to the continuum limit using the renormalization
equations (\ref{equ:latren}), and in this way
both phases have been found also
in the continuum theory.
The phase diagram is shown in Fig.~\ref{fig:phasediag}.
There is a first-order transition
at small $x$, as expected from the perturbative approximation 
\cite{ref:halperin}. At large $x$, the transition seems to be
continuous and its order is not known. In any case, the latent heat of
the first-order transition is too small to be observed experimentally
in superconductors.

\section{Vortices}
The two-dimensional XY model is a well-known example of a theory
that displays a phase transition without a symmetry breaking
\cite{ref:kosterlitz}. In the 
low-temperature phase, the
dominating excitations are the spin waves,
and the correlation length of the system is infinite. Fluctuations
create small vortex-antivortex pairs, but they do
not have much effect on any large-scale properties. 
At high enough temperature,
nevertheless, the fluctuations are capable of creating vortices and
antivortices arbitrarily far apart from each other and therefore
they become
the dominating excitations. This makes the correlation length finite.
Therefore, there must be a phase transition point at which the
correlation length diverges.

The Ginzburg-Landau theory (\ref{equ:contlagr}) has similar properties.
It has two phases, one of which has an infinite correlation length.
It also contains vortices,
in the form of closed loops
\cite{ref:nielsenolesen}.
To see this, consider a field configuration $(\psi(\x),\A(\x))$.
Let us write $\psi(\x)=\rho(\x)\exp[\I\gamma(\x)]$, where $\gamma$ is
only defined modulo $2\pi$ and has singularities when $\psi(\x)=0$.
If we require that $\gamma$ is continuous when $\psi(x)\neq0 $, it becomes a 
multi-valued function. Therefore it is possible that the contour integral
around a closed curve $C$ 
\beq
\label{equ:jatkumojanne}
\oint_C \dd\x\cdot\nab\gamma(\x)\equiv 2\pi n_C
\eeq
can have non-zero values. The winding number $n_C$ is a
gauge-invariant integer and can be non-zero if there is
a singularity, a vortex line, going through the curve.
These vortex lines cost energy and they can be removed only
by shrinking them to a point.
Therefore they are
classically stable objects. It is natural to think that
in the superconducting phase, the fluctuations can create only
small vortex loops, but in the normal phase, they can be arbitrarily
large. However, the problem is how to calculate their distribution,
or the expectation value of $|n_C|$, to confirm this idea.

The effects of the vortices have been studied intensively in the
London limit $\lambda\rightarrow\infty$ of the lattice theory 
\cite{ref:vortexpapers}. 
In this limit,
the length of $\psi$ is fixed and the theory
can be further simplified by replacing the Lagrangian with
the Villain version \cite{ref:Villain}. One can then construct
a dual theory \cite{ref:Chavel} in which the vortex loops
are the fundamental objects. The phase transition takes place
because of condensation of vortices. It still remains to be seen,
to what extent the same is true in the realistic case, i.e.~when 
$a\rightarrow 0$ together with $\lambda\rightarrow 0$.

To answer this question, Monte Carlo simulations must be performed
in the original lattice theory (\ref{equ:latlagr}). Then
the relations (\ref{equ:latren}) make the continuum extrapolation
of physical quantities possible.
To study the properties of the vortices,
the winding number (\ref{equ:jatkumojanne}) must be also defined
on the lattice, but the most straightforward choice, i.e. replacing
the gradient by its lattice counterpart
\beq
\label{equ:latticegrad}
\partial_i\gamma(\x)\rightarrow\frac{1}{a}[\gamma(\x+\hat\imath)-\gamma(\x)]_\pi
\eeq
leads to problems. The resulting winding number would not be gauge
invariant, since one can, for example, choose $\theta(\x)=-\gamma(\x)$
in Eq.~(\ref{equ:gaugelat}), in which case there would be no vortices.
One solution would be to fix the gauge, but then the result would
depend on the gauge chosen and would be difficult to interpret.
Therefore a gauge-invariant lattice analogue for Eq.~(\ref{equ:jatkumojanne})
is needed.

For each positively directed link $l=(\x,\x+\hat\imath)$, let us define
\beq
\label{equ:def1}
Y_l=
[\alpha_i(\x)+\gamma(\x+\hat\imath)-\gamma(\x)]_\pi-\alpha_i(\x).
\eeq
This is clearly nothing but
\beq
\label{equ:def2}
Y_l=\gamma(\x+\hat\imath)-\gamma(\x)+2\pi n_i(\x),
\eeq
where $n_i(\x)$ is such an integer that 
$Y_l\in(-\pi-\alpha_i(\x),\pi-\alpha_i(\x)]$.
For links with negative direction, $\l'=(\x+\hat\imath,\x)$, 
we define the sign to be the opposite:
$
Y_{l'}=-Y_l.
$
Then, for each closed loop $C$, we can define the winding number $n_C$ as
\beq
\label{equ:correct}
Y_C=\sum_{l\in C}Y_l\equiv 2\pi n_C.
\eeq
This definition of the winding number coincides with that given in 
Ref.~\cite{ref:Chavel} for the $\lambda=\infty$ case.

From Eq.~(\ref{equ:def2}), it is easy to see that $n_C$ is an integer, 
since in the sum every
$\gamma$ appears twice with opposite signs and only $n_i$ give
a non-vanishing contribution.
It is also gauge invariant, since
the part of Eq.~(\ref{equ:def1}) in the brackets is invariant by itself,
and the contribution from the last term $-\alpha_i(\x)$ to 
Eq.~(\ref{equ:correct}) gives only the magnetic flux through the curve $C$,
which is a gauge-invariant quantity. 
The winding number has also the
correct continuum limit (\ref{equ:jatkumojanne}) in the sense that
$\alpha_i$ in Eq.~(\ref{equ:def1}) is proportional to the
lattice spacing and what remains is exactly the gradient of the
phase $\gamma$.

The winding number $n_C$ is additive in the sense that if $C$ is 
composed of two curves
$A$ and $B$, $n_C=n_A+n_B$. This is of course necessary for it to be
meaningful to think of $n_C$ as the number of vortices going through $C$.
It also means that vortices do not end,
but form closed loops. Therefore Eq.~(\ref{equ:correct}) is a valid definition
for the existence of a vortex on a lattice.

\begin{figure}
\begin{center}
\epsfig{file=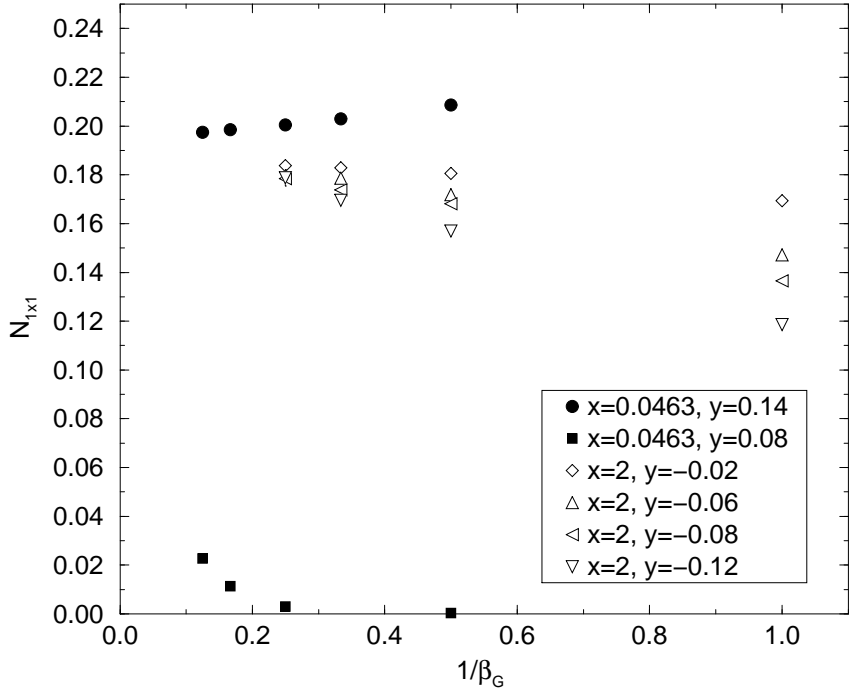,width=13cm}
\end{center}
\caption{
\label{fig:n1x1}
The values of $N_{1\times 1}\equiv\langle n_{1\times 1}\rangle$ 
at various lattice spacings
$a=1/\beta_G$ measured at the points shown in Fig.~\ref{fig:phasediag}
\cite{ref:PRL}.
Far from the continuum limit, a discontinuity is found at small $x$, but
not at large $x$. Note, however, that the values converge to the same
point on the continuum limit, and no dependence on the parameters or
the phase of the system remains.
}
\end{figure}

Let us then consider the Monte Carlo simulations \cite{ref:PRL}.
The simplest quantity to measure is the expectation value of number 
of vortices through a single 
plaquette, $\langle |n_{1\times 1}| \rangle$. The absolute value has to
be taken since otherwise the contribution of 
vortices going through the plaquette in
opposite directions would cancel each other and the result would be zero.
This is a non-trivial quantity at finite lattice spacing and it behaves
as is expected for the vortex density of the system: It is large in
the normal and small in the superconducting phase (See Fig.~\ref{fig:n1x1}). 
Nevertheless, it is
not a true order parameter since it does not vanish in the superconducting
phase. 

To find an observable that really can distinguish between the two phases,
we have to consider non-local ones.
One qualitative difference in the vortex distributions
above and below the critical temperature is expected to be
the presence of percolating
vortices in the high-temperature phase. 
There is percolation in the
system if a vortex that
extends through the whole lattice
is found with a non-zero probability even in the infinite-volume limit.
To observe this, one has to trace the individual vortices on the lattice.
Of course, it is possible that the percolation and the true critical point
do not coincide, and numerical simulations cannot exclude this possibility,
but they can still give much insight to the problem.

Although the lattice Ginzburg-Landau model (\ref{equ:latlagr})
has interesting and non-trivial properties even at non-zero $a$, 
the continuum limit 
$a\rightarrow 0$ has
more physical significance. Therefore one should ask, which quantities
have a finite continuum limit. Unlike in four-dimensional field theories,
there is no problem with the renormalization of the theory, since
the relations (\ref{equ:latren}) allow one to express everything in terms
of renormalized continuum quantities. 
Since $\langle|n_{1\times 1}|\rangle$ depends only on the values of the fields
in a finite-size region in lattice units, it measures only the
ultraviolet properties. Actually, at the continuum limit 
its value coincides with the
analogous quantity in
the massless free-field theory and is independent of the
parameters $x$ and $y$ and the phase of the system (See Fig.~\ref{fig:n1x1}).
Therefore it is not an interesting quantity from our point of view.

To find non-trivial quantities we need to
consider observables that measure the physics at finite distances
in physical units. 
The choice of Ref.~\cite{ref:PRL} was to increase the size of 
the curve $C$ in $\langle|n_C|\rangle$
accordingly in such a
way that for each value of $a$ it is of the same size in physical
units. An example is a square of $(c/a)\times (c/a)$ plaquettes.
However, one has to remove the contribution of the ultraviolet
effects to get physically meaningful results. 
It is not clear whether there is a way to do this for $\langle|n_C|\rangle$
itself,
but in Ref.~\cite{ref:PRL} evidence was given for its discontinuity 
at the phase transition line 
to be independent of the ultraviolet details.

In subsequent studies the emphasis should be on the spatial distribution
of the vortices. Suitable observables for that purpose are
correlation lengths of the vortex density. If we denote 
by $n_{ij}(\x)$ the winding number of the plaquette 
$(\x,\x+\hat\imath,\x+\hat\imath+\hat\jmath,\x+\hat\jmath)$,
the correlation lengths are given by 
the exponential decay of the correlators
\beq
G_{ij,kl}(\x-\y)=\langle n_{ij}(\x) n_{kl}(\y)\rangle.
\eeq
Percolation can also be studied on the continuum limit but
it may be numerically demanding, since at small $a$ there will
be many vortices present also in the superconducting phase.

\section{Conclusions}
We have discussed the properties of the phase transition of the
Ginzburg-Landau theory. To stress the importance of the fluctuations
in this problem, we gave a brief review of
the phenomenon of spontaneous symmetry 
breakdown and Elitzur's theorem. We pointed out that, like in the XY model,
vortices may have an important role in the transition. 
The only systematic way to study the behavior of the vortices are
lattice simulations, but
the naive discretization of the continuum winding number does not give
a valid lattice observable since it is not gauge-invariant and leads
to trivial results. Therefore we defined 
a gauge-invariant criterion for the existence
of vortices. We also suggested some vortex-related 
quantities that would give interesting
information on the phase transition. The full details and the
analysis of the results of the Monte Carlo simulations are
given in Ref.~\cite{ref:PRL}.

In the studies of vortices in the superconductors, one is usually
interested in their behavior in an external magnetic field. The formalism
discussed here can also be extended to that case. 
In cosmology, the interesting question is the nature and the subsequent
evolution of the vortex
network created in a phase transition. 
Our approach
can be used for that purpose by calculating the properties of
the network at the instant when the strings fall out of equilibrium.
This gives the initial conditions for the time evolution of the strings.
As opposed to the mean-field approach, any value of the parameter $x$
can be used.

\begin{ack}
I would like to thank the organizers of the symposium
for a successful meeting, 
and G.~Blatter, T.W.B.~Kibble and A.~Kupiainen for useful
discussions. I also thank K.~Kajantie, M.~Karjalainen, M.~Laine and
J.~Peisa for collaboration on this topic.

The work was partially supported by the TMR network
{\em Finite Temperature Phase Transitions in Particle Physics},
EU contract no. ERBFMRXCT97-0122.
\end{ack}

\end{document}